\documentclass[review,a4paper,3p,12pt]{elsarticle}

\usepackage[T1]{fontenc}
\usepackage[utf8]{inputenc}
\usepackage{ams math,amssymb}
\usepackage[colorlinks=true,citecolor=ForestGreen,linkcolor=Red,urlcolor=Blue,hypertexnames=false]{hyperref}
\usepackage{chngcntr}

\usepackage{graphicx}

\usepackage{natbib}
\usepackage[dvipsnames]{xcolor}
\usepackage{hyperref}
\usepackage{booktabs}
\usepackage{color}
\counterwithout{figure}{section}

\graphicspath{{Images/}}

\begin{document}

\begin{frontmatter}
\title{Group cohesion under asymmetric voting behaviors}

\author[1]{Hao Yu}
\author[1]{Youjin Wen}
\author[1]{Zhehang Xu}
\author[1]{Jianlin Zhang}

\author[1,5]{Fanyuan Meng}

\address[1]{Research Center for Complexity Sciences, Hangzhou Normal University, Hangzhou,311121, Zhejiang, China}



\address[5]{Corresponding authors: fanyuan.meng@hotmail.com}

\begin{abstract}

Cohesion plays a crucial role in achieving collective goals, promoting cooperation and trust, and improving efficiency within social groups. To gain deeper insights into the dynamics of group cohesion, we have extended our previous model of noisy group formation by incorporating asymmetric voting behaviors. Through a combination of theoretical analysis and numerical simulations, we have explored the impact of asymmetric voting noise, the attention decay rate, voter selection methods, and group sizes on group cohesion. For a single voter, we discovered that as the group size approaches infinity, group cohesion converges to $1/(R+1)$, where $R$ represents the ratio of asymmetric voting noise. Remarkably, even in scenarios with extreme voting asymmetry ($R \to \infty$), a significant level of group cohesion can be maintained. Furthermore, when the positive or negative voter's voting noise surpasses or falls below the phase transition point of $R_c=1$, a higher rate of attention decay can lead to increased group cohesion. In the case of multiple voters, a similar phenomenon arises when the attention decay rate reaches a critical point. These insights provide practical implications for fostering effective collaboration and teamwork within growing groups striving to achieve shared objectives.

\end{abstract}


\end{frontmatter}

\section{Introduction}
Social groups are composed of individuals who share common goals or interests and possess a sense of community \cite{bruggeman2017solving,stangor2015social}. These groups emerge naturally within society and serve crucial functions in areas such as innovation \cite{young2011dynamics,kim2015impacts,nijstad2002creativity}, communication \cite{huckfeldt1995political,o1977task,chockler2001group}, and marketing \cite{smith2006social,wang2013discovering}. Therefore, investigating the formation and evolution of social groups enhances our comprehension of the underlying mechanisms that contribute to social cohesion and stability.

Group cohesion is a widely researched concept in sociology and psychology \cite{evans1980group,webber2001impact,yoo2001media}, and pertains to the level of unity and solidarity within a group. Understanding group cohesion is crucial for comprehending group dynamics and behavior. This is especially true for teams in which consensus is vital for achieving success \cite{dobbins1986effects,friedkin2004social}. Numerous studies consistently demonstrate that cohesive groups tend to perform better in their tasks \cite{carron2002team,casey2009sticking}. 

Group cohesion is influenced by various factors, one of which is group attraction, closely related to the acceptance of new members \cite{zander1979psychology,hogg1993group}. However, empirical studies have indicated that larger groups often exhibit lower levels of cohesion 
 \cite{snijders2011statistical}. While attracting new members can bring benefits to a group, it also poses the risk of reduced cohesion, leading to group fragmentation, and instability \cite{meng2023disagreement,fenoaltea2023phase}. To illustrate the significant implications of reduced cohesion, we can examine the Partition of India—an epoch-making event that occurred after the independence of British India in 1947. Driven by religious and political differences between Hindus and Muslims, it led to the division of the region into two separate nations, India and Pakistan, and triggered large-scale population displacements and enduring conflicts. This division and its associated strife had profound consequences not only for India and Pakistan but also for the wider South Asian region. It gave rise to persistent enmity, complex security dynamics, and a detrimental cycle, severely undermining regional stability, development, and cooperation \cite{talbot2009partition,ambedkar1945pakistan,brass2003partition}.

To mitigate the adverse consequences associated with the decline of group cohesion, it is essential to conduct a comprehensive investigation into the factors influencing its dynamics. In this context, cohesion refers to the dynamic process that reflects a group's tendency to unite and maintain unity \cite{brass2003partition}. Expanding on our previous research \cite{PhysRevResearch.5.013023}, we aim to delve into the effects of asymmetric voting behaviors on group cohesion. Our study incorporates non-Markovian properties of asymmetric voting noise, wherein the voting noise can increase due to individuals' contributed admissions. We comprehensively investigate the impact of asymmetric voting noise, different attention decay rates, two distinct methods for selecting voters, and varying group sizes on group cohesion. Our research findings demonstrate that in an infinitely large group, group cohesion tends to converge to a constant value determined by the ratio of asymmetric voting noise. Moreover, increased attention decay does not always have a detrimental effect on group cohesion.


\section{Model}
We consider a scenario in which a group of individuals holds two types of opinions, denoted as $o \in \{-1, +1\}$. At time step $t = 0$, the group consists of $N_0$ individuals, all sharing the opinion $+1$. As time progresses ($t = 1, 2, \dots$), a candidate $j$ is randomly chosen with an equal probability of having the opinion $o_j = +1$ or $o_j=-1$. From the existing group members, a voter $i$ with opinion $o_i$ is selected based on a specific rule. If the selected voter has an opinion $o_i = +1$, the candidate with opinion $o_j = +1$ or $o_j = -1$ can join the group with probabilities of $1-\alpha$ or $\alpha$, respectively, through positive voting. Conversely, if the chosen voter has an opinion $o_i = -1$, the candidate with opinion $o_j = -1$ or $o_j = +1$ can join the group with probabilities of $1-\beta$ or $\beta$, respectively, through positive voting. Candidates who fail to be accepted through positive voting are rejected through negative voting. The parameters $\alpha \in [0,0.5]$ and $\beta \in [0,0.5]$ reflect the irrational behavior of the voters with different opinions and can be regarded as noise parameters. Unequal probabilities $\alpha$ and $\beta$ indicate asymmetric voting behavior within the group. We define the ratio of asymmetric voting noise as $R = {\alpha}/{\beta} \in [0, \infty]$.

We examine two distinct methods of selecting voters:

1) Uniform selection (UC): Voters are selected uniformly at random from the existing group members.

2) Preferential attachment (PA): Voters are selected proportionally to their past contributions in terms of admissions. This means that members who have made more admissions in the past have a higher likelihood of being chosen as voters.

The group's expansion continues until a predetermined group size, denoted as $N$, is reached. Afterward, we assess the resulting cohesion $C$ of the group using the following formula
\begin{equation}
    C = \frac{|\{i:o_i=+1\}_{i=1}^N|}{N},
\end{equation}

where $|\{i:o_i=+1\}_{i=1}^N|$ represents the count of members holding the opinion $+1$. The cohesion $C$ represents the proportion of members holding the opinion $+1$ within the group.

On the one hand, when $\alpha = 0$, it means that candidates with the opinion $-1$ always receive negative voting and are therefore rejected from the group. In other words, since no members with the opinion $-1$ are allowed to join the group, then the group cohesion is always 1. On the other hand, when $R=1$, as the group size $N$ approaches $\infty$, the group cohesion tends to 0.5.

\section{Results}\label{sec:result}
\subsection{Preliminaries: Symmetric voting noise}
To examine the impact of asymmetry, we can first consider the case where $\alpha=\beta$ (i.e. $R=1$), which serves as a natural benchmark. When $\alpha=\beta$, the model reduces to the one presented in \cite{PhysRevResearch.5.013023}. Let's define the probability that a candidate is allowed into the group as $P(t)$. Based on the arguments in \cite{PhysRevResearch.5.013023}, we can derive the following recursive equation for $P(t)$ with the initial condition $P(0)=1$ as 

\begin{equation}
    P(t+1) = \left\{
    \begin{aligned}
    & \Big(1-\frac{2\alpha}{t+N_0}\Big)P(t)+\frac{\alpha}{t+N_0} \quad \text{(UC)}, \\
    & \Big(1-\frac{2\alpha}{2t+N_0(N_0-1)}\Big)P(t)+\frac{\alpha}{2t+N_0(N_0-1)} \quad \text{(PA)}.
    \end{aligned}
    \right.
\end{equation}

Taking the average of the aforementioned solution for $P(t)$ over $t=0,\dots,N-N_0$, we obtain the mean cohesion for large group size as follows
\begin{equation}
    \overline{C(N,N_0,\alpha)} \approx \left\{
    \begin{aligned}
    & \frac{1}{2} +
\frac{\Gamma(N_0+1)}{2\Gamma(N_0+1-2\alpha)}\,N^{-2\alpha} \quad \text{(UC)}, \\
    & \frac{1}{2} +
\frac{(1-2\alpha)\Gamma(N_0(N_0-1)/2+1)}{2(1-\alpha)\Gamma(N_0(N_0-1)/2+1-\alpha)}\,N^{-\alpha} \quad \text{(PA)}.
    \end{aligned}
    \right.
\end{equation}

\subsection{Asymmetric voting noise}
Whenever $\alpha \neq \beta$ (i.e. $R\neq 1$), we refer to the presence of asymmetric voting behaviors accompanied by asymmetric voting noise. Let's first examine the case where the voter is chosen randomly (UC). We can determine the probability $W(i)$ that voter $i$ in the group gives a positive voting (+1) using the following expression

\begin{equation}
    W(i) = P(i)(1 - \alpha) + (1 - P(i))\beta.
\end{equation}

To obtain the probability $P(t)$, we average over all the $W(i)$ values for $i<t$. Therefore, we have

\begin{equation} \label{eq:master_uc}
    P(t+1) = \frac{N_0}{t+N_0}W(0) + \frac{1}{t+N_0}\sum_{i=1}^t W(i).
\end{equation}

After some algebraic manipulation, we can rearrange Eq.\ref{eq:master_uc} as

\begin{equation}
    P(t+1) = \left(1 - \frac{\alpha+\beta}{t+N_0}\right)P(t) + \frac{\beta}{t+N_0}.
\end{equation}

By averaging the solution over $t = 0, . . . , N - N_0$, we obtain the mean cohesion

\begin{equation} \label{eq:cohesion_uc}
    \overline{C(N,N_0,\alpha,\beta)_{UC}} = \frac{1}{R+1}  + \frac{R}{R+1}\frac{\Gamma(N_0)\Gamma(N-\alpha - \beta)}{\Gamma(N_0-\alpha - \beta)\Gamma(N)}.
\end{equation}

When $N_0$ is fixed and $N \gg N_0$, Eq.\ref{eq:cohesion_uc} implies

\begin{equation} \label{eq:cohesion_n_infinity}
    \overline{C(N,N_0,\alpha,\beta)_{UC}} \approx \frac{1}{R+1} + \frac{R}{R+1} \frac{\Gamma(N_0)}{\Gamma(N_0-\alpha-\beta)}N^{-\alpha-\beta}.
\end{equation}

Now let's examine the scenario where the probability of selecting a voter is proportional to the number of contributed admissions (PA). Following the arguments presented in \cite{PhysRevResearch.5.013023}, we can express the dynamics of the number of contributed admissions, denoted as $k_i(t)$, for individual $i$ as follows

\begin{equation}
    k_i(t+1) = k_i(t) + \frac{k_i(t)}{2t+N_0(N_0-1)}.
\end{equation}

Thus the recursive form for $P(t)$ becomes

\begin{equation} 
    P(t+1) = \left(1 - \frac{\alpha+\beta}{2t+N_0(N_0-1)}\right)P(t) + \frac{\beta}{2t+N_0(N_0-1)}.
\end{equation}

The mean cohesion can then be obtained as follows

\begin{equation} \label{eq:cohesion_pa}
    \overline{C(N,N_0,\alpha,\beta)_{PA}} = \frac{1}{R+1} + \frac{R}{R+1} \frac{\Gamma(\frac{N_0(N_0-1)}{2})\Gamma(\frac{N_0(N_0-1)-\alpha-\beta}{2}+N-N_0)}{\Gamma(\frac{N_0(N_0-1)-\alpha-\beta}{2})\Gamma(\frac{N_0(N_0-1)}{2}+N-N_0)}.
\end{equation}

When $N_0$ is fixed and $N \gg N_0$, Eq.\ref{eq:cohesion_pa} implies

\begin{equation}
    \overline{C(N,N_0,\alpha,\beta)_{PA}} \approx \frac{1}{R+1} + \frac{R}{R+1} \frac{\Gamma(\frac{N_0(N_0-1)}{2})}{\Gamma(\frac{N_0(N_0-1)-\alpha-\beta}{2})} N^{\frac{-\alpha-\beta}{2}}.
\end{equation}

\begin{figure}[htb] 
    \centering
    \includegraphics[width=\linewidth]{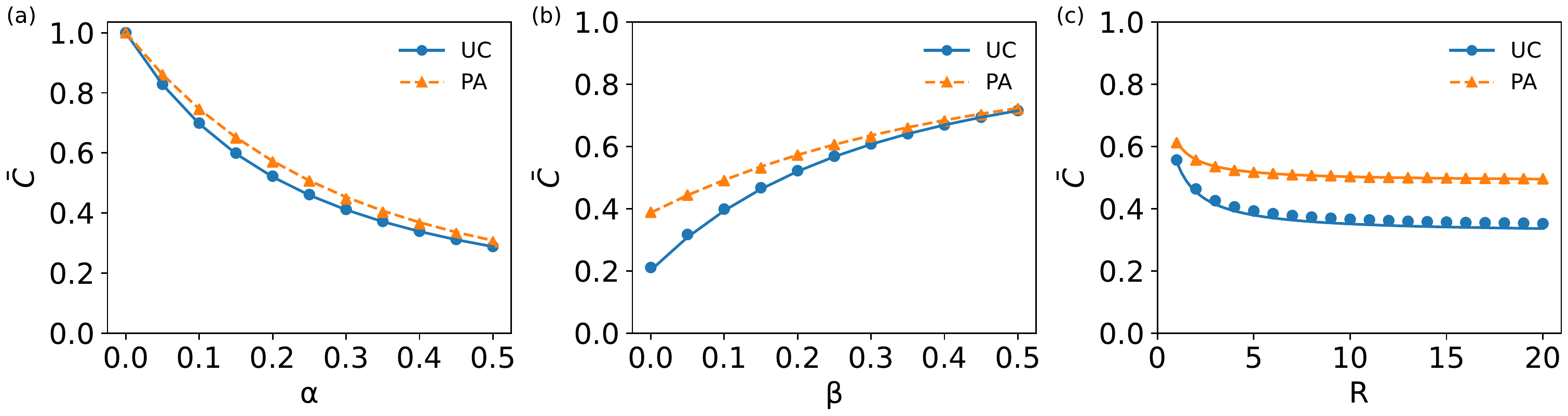}
     \caption{
    The relationship between mean cohesion $\overline{C}$ and the asymmetric voting noise $\alpha, \beta$ is shown for various selection methods. The markers represent simulation results, while the dashed lines represent theoretical predictions with $N_0 = 4$ and $N = 10^{4}$. The three panels depict: (a) $\beta=0.2$, (b) $\alpha=0.2$, and (c) $\alpha=0.2$ with $R$ varying from 1 to 20.}
    \label{fig:c_noise}
\end{figure}

Eqs.\ref{eq:cohesion_uc} and \ref{eq:cohesion_pa} imply that when $\beta$ is fixed (e.g. $\beta=0.2$), increasing $\alpha$ leads to a decrease in the mean cohesion $\overline{C}$. Additionally, the mean cohesion in the PA case is always greater than or equal to that in the UC case (see Fig.\ref{fig:c_noise}(a)). Similarly, when $\alpha$ is fixed (e.g. $\alpha=0.2$), increasing $\beta$ results in an increase in the mean cohesion $\overline{C}$. Again, the mean cohesion in the PA case is always greater than or equal to that in the UC case (see Fig.\ref{fig:c_noise}(b)). In a special case where $\alpha=0$ and $\beta\neq0$ (i.e. $R=0$), the cohesion is always equal to 1 regardless of the group size. However, in another special case where $\alpha \neq 0$ and $\beta=0$ (i.e. $R \to \infty$), Eqs.\ref{eq:cohesion_uc} and \ref{eq:cohesion_pa} can be rewritten as 

\begin{equation}
    \overline{C(N,N_0,\alpha,\beta)}  = \left\{
    \begin{aligned}
    & \frac{\Gamma(N_0)\Gamma(N-\alpha)}{\Gamma(N_0-\alpha)\Gamma(N)} \quad \text{(UC)}, \\
    & \frac{\Gamma(\frac{N_0(N_0-1)}{2})\Gamma(\frac{N_0(N_0-1)-\alpha}{2}+N-N_0)}{\Gamma(\frac{N_0(N_0-1)-\alpha}{2})\Gamma(\frac{N_0(N_0-1)}{2}+N-N_0)} \quad \text{(PA)}, 
    \end{aligned}
    \right.
\end{equation}
which implies that as the ratio of asymmetric voting noise $R$ tends to infinity, the value of mean cohesion will converge for both the UC and the PA cases (see Fig.\ref{fig:c_noise}(c)).

\subsection{Asymmetric voting noise with attention decay}
To model the dynamic and non-linear evolution of attention decay for voters, we introduce non-Markovian properties for noise parameters in our model. Thus we define the following equations

\begin{equation} \label{eq:decay}
\left \{
 \begin{aligned}
        \alpha_i(t) = \alpha + \left(\frac{1}{2} - \alpha\right)\tanh \Big[b\Big(k_i(t)-1\Big)\Big], \\
            \beta_i(t) = \beta + \left(\frac{1}{2} - \beta\right)\tanh \Big[b\Big(k_i(t)-1\Big)\Big].
 \end{aligned}
 \right.
\end{equation}

Here, $\tanh$ is the hyperbolic tangent function, and the attention decay rate is denoted as $b \geq 0$. Therefore, $\alpha_i(t) \in [\alpha, 0.5]$ and $\beta_i(t) \in [\beta, 0.5]$ represent the voting noise for a voter $i$ with opinion $+1$ and opinion $-1$ at time $t$, respectively.

In the UC case with the infinite group size (i.e. $N \to \infty$), we can apply the mean-field method to derive a simplified expression for Eq.\ref{eq:decay} as follows

\begin{equation}
\left \{
 \begin{aligned}
         \langle \alpha \rangle = \alpha + \left(\frac{1}{2} - \alpha\right)\tanh \Big[b\left(\langle k \rangle -1\right)\Big], \\
              \langle \beta \rangle = \beta + \Big(\frac{1}{2} - \alpha\Big)\tanh \Big[b\Big(\langle k \rangle -1\Big)\Big].
 \end{aligned}
 \right.
\end{equation}

Thus, the mean cohesion for the infinite group size can be expressed as

\begin{equation}
    \overline{C} (\infty) \Big(1-\langle \alpha \rangle \Big) + \Big(1 - \overline{C}(\infty)\Big) \langle \beta \rangle = \overline{C}(\infty).
\end{equation}

Finally, we obtain the expression for mean cohesion

\begin{equation} \label{eq:cohesion_non_infinity}
    \overline{C}(\infty) = \frac{\beta + (1/2-\beta) \tanh \Big[b\Big(\langle k \rangle -1\Big)\Big]}{\alpha + \beta + (1-\alpha - \beta) \tanh \Big[b\Big(\langle k \rangle -1\Big)\Big]}.
\end{equation}

\begin{figure}[htb]
    \centering
    \includegraphics[width=\linewidth]{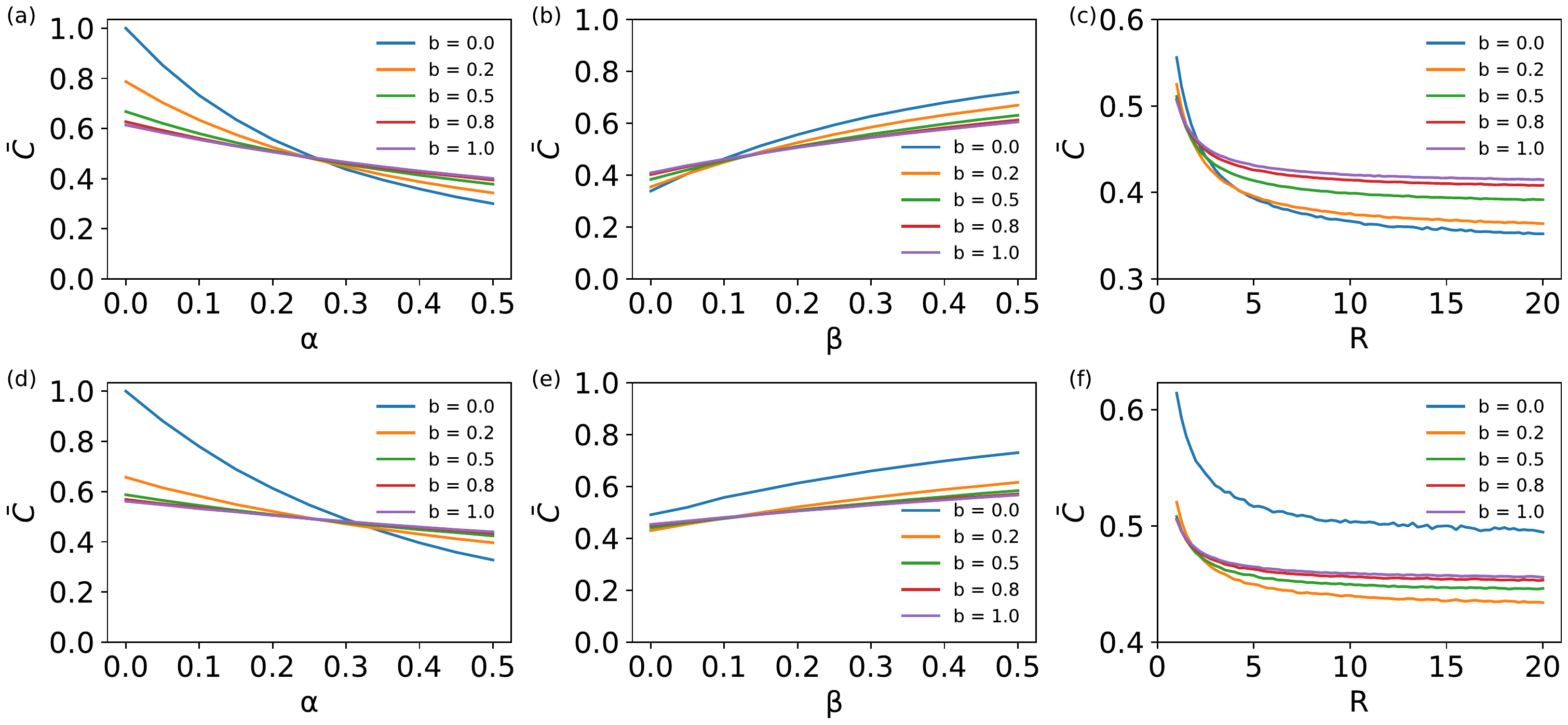}
    \caption{The simulation results depict the relationship between mean cohesion $\overline{C}$ and voting noises $\alpha$, $\beta$, as well as the ratio of asymmetric voting noise $R$, for different values of $b$. Panels (A-C) illustrate the UC case with fixed values of $\beta = 0.2$ and $\alpha = 0.2$, while $R$ varies from 1 to 20, respectively. Panels (D-F) correspond to the PA case with fixed values of $\beta = 0.2$ and $\alpha = 0.2$, while $R$ varies from 1 to 20, respectively. The group size is set to $N=10^3$.}
    \label{fig:cohesion_uc_non}
\end{figure}

We can therefore make an interesting discovery regarding the relationship between mean cohesion $\overline{C}(\infty)$ and voting noise for both the UC case regardless of the values of $b$. Specifically, we observe the emergence of critical point $R_c=1$ with $\overline{C}=0.5$ (confirmed by Eq.\ref{eq:cohesion_non_infinity}) as we vary the value of $b$ (see Fig.\ref{fig:cohesion_uc_non}(a-c)). This intriguing result suggests that the decay of attention does not always have a negative impact on group cohesion. In other words, a higher rate of attention decay can lead to increased group cohesion. 

Additionally, it is worth noting that the overall decreasing or increasing tendencies of the relationship between mean cohesion and voting noise remain consistent, although the specific values may vary for different $b$ values. Moreover, as the ratio of asymmetric voting noise $R$ approaches infinity, the mean cohesion value converges for both the UC and PA cases (refer to Fig.\ref{fig:cohesion_uc_non}). Furthermore, our analysis reveals that the non-Markovian decay of noises has a more significant effect on mean cohesion in the UC case compared to the PA case, as depicted in Fig.\ref{fig:cohesion_uc_non}.


Further, we have conducted an extensive investigation into the relationship between mean cohesion $\overline{C}$ and the attention decay rate $b$ for different numbers $m$ of voters at each time step. The candidate can be admitted to the group only if all voters provide positive votes. Our findings reveal the existence of a critical point $b_c$ in both the UC and PA cases. When the attention decay rate $b$ exceeds the critical value of $b_c$, the mean cohesion transitions from a decreasing trend to an increasing one. 

\begin{figure}[htb]
    \centering
    \includegraphics[width=\linewidth]{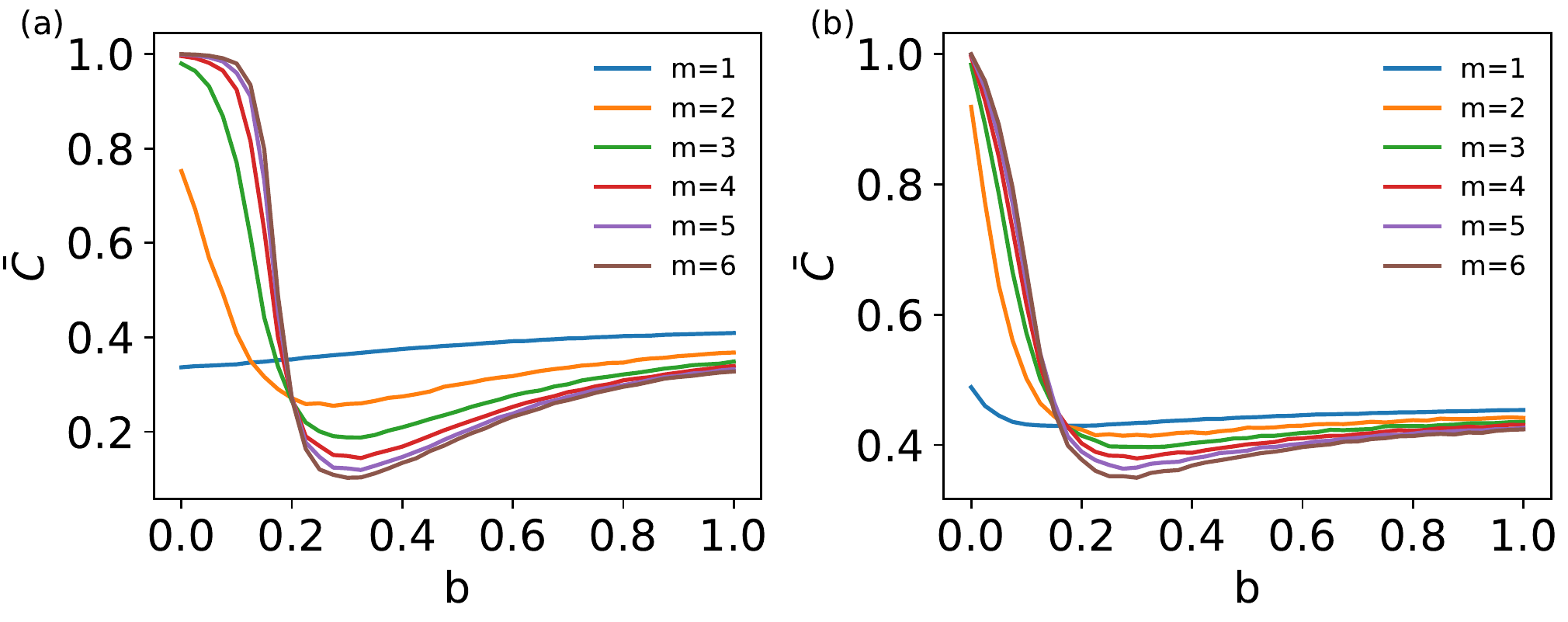}
    \caption{
The relationship between mean cohesion $\overline{C}$ and attention decay rate $b$ is depicted in the two panels, corresponding to the UC (a) and PA (b) cases. The simulations were conducted with different numbers $m$ of voters. In both panels, the parameters were set as follows: $\alpha=0.2$, $\beta=0$, $N_0 = m$, and $N = 10^3$.}
    \label{fig:b_m}
\end{figure}

In particular, for the UC case, we observe that as $b$ increases, all curves intersect at approximately $b \approx 0.2$, except for the case of $m=1$ (refer to Fig.\ref{fig:b_m}(a)). On the other hand, in the PA case, as $b$ increases, all curves intersect at a point around $b \approx 0.18$ (refer to Fig.\ref{fig:b_m}(b)). These results provide valuable insights into the interplay between attention decay and mean cohesion, demonstrating the critical role of the attention decay rate in shaping the dynamics of group cohesion.

\newpage
\section{Conclusion and Discussion}

In conclusion, we have proposed a model of noisy group formation by incorporating asymmetric voting behaviors. Our work complements and extends the original model proposed in \cite{meng2023disagreement}. By departing from the unrealistic assumption of homogeneous judgment capabilities among individuals of different types and considering asymmetric voting noise with attention decay, we provide a more realistic representation of group dynamics. Through a combination of theoretical analysis and numerical simulations, we have explored the impact of asymmetric voting noise, attention decay, voter selection methods, and group sizes on group cohesion.


Furthermore, our model opens up several potential extensions and research directions. It can be expanded to encompass multi-dimensional scenarios where the opinions of individuals can influence each other, utilizing matrices to simulate interactions. The possibility of member dissatisfaction and the dynamics of individuals leaving a group due to significant differences can be incorporated using evolutionary game theory. Additionally, incorporating cost-benefit analysis and considering opinions changing due to peer influence would provide a more comprehensive understanding of group dynamics.

In summary, our work provides a novel framework and approach for understanding the formation and evolution of social groups. It sheds light on the effects of different voting behaviors, and group size on group cohesion, as well as the spontaneous decrease and phase transition of group cohesion. These findings offer valuable insights into the dynamic mechanisms and social implications of group cohesion, contributing to a deeper understanding of team dynamics and facilitating effective collaboration within groups striving to achieve common objectives.

\bibliographystyle{unsrt}

\bibliography{main.bib}

\end{document}